\documentstyle[prl,aps,preprint]{revtex}
\begin{document}
\tighten

\title{THE SPIN STRUCTURE OF THE NUCLEON \\ IN THE ASYMPTOTIC LIMIT
\thanks {This work is supported in part by funds provided by the
U.S.~Department of Energy (D.O.E.) under cooperative agreement
\#DF-FC02-94ER40818
and by the
National Science Foundation grant INT--9122027
(US--Pakistan Collaboration on Hadron Structure at High Energies).
}}

\author{Xiangdong Ji and Jian Tang \\[0.7ex]}

\address{Center for Theoretical Physics, \\
Laboratory for Nuclear Science and Department of Physics \\
Massachusetts Institute of Technology, \\
Cambridge, Massachusetts 02139}

\author{Pervez Hoodbhoy \\[0.7ex]}

\address{Center for Theoretical Physics, \\
Laboratory for Nuclear Science and Department of Physics \\
Massachusetts Institute of Technology, \\
Cambridge, Massachusetts 02139 \\
{\it and} \\
Physics Department \\
Quaid-e-Azam University \\
Islamabad, Pakistan{~}\\[1.4ex]}

\date{MIT-CTP-2476 {~~~} hep-ph/9510304 {~~~} October 1995}

\maketitle

\begin{abstract}

\vspace*{-6.0ex}

In analogy to the Altarelli-Parisi equation for the quark and gluon
helicity contributions to the nucleon spin, we derive an evolution
equation for the quark and gluon orbital angular momenta.  The solution
of the combined equations yields the asymptotic fractions of the nucleon
spin carried by quarks and gluons: $3n_f/(16+3n_f)$ and $16/(16+3n_f)$,
respectively, where $n_f$ is the number of active quark flavors.  These
are identical to the well-known asymptotic partitions of the nucleon
momentum between quark and gluon contributions.  We show that the
axial-anomaly contribution to the quark helicity is cancelled by a
similar contribution to the quark orbital angular momentum, making the
total quark contribution to the nucleon spin anomaly-free.

\end{abstract}
\pacs{xxxxxx}

\narrowtext

What are the composition of the nucleon spin
in terms of its quark and gluon
constituents? In the last few years, the answer
has become the holy grail for the nuclear and particle
spin community. The motivation is quite clear: given Quantum
Chromodynamics (QCD) is difficult to solve, such information
gives us valuable insights into the nonperturbative wavefunction
of the nucleon. A satisfactory understanding of the
spin structure would be to know, for instance, how much
spin of the nucleon is carried, respectively, by the quark
and gluon helicities and orbital angular momenta, in the same
way as one now understands how the mass of the
nucleon is partitioned among contributions
from the quark and gluon kinetic energies,
quark masses, and the trace anomaly\cite{ji}.

The theoretical basis for separating the spin of the
nucleon into different contributions begins with the QCD expression
for the generators of Lorentz transformations \cite{jaffemanohar},
\begin{equation}
     J^{\mu\nu} = \int d^3 x M^{0\mu\nu} \ ,
\end{equation}
where $M^{\alpha\mu\nu}$ is the angular momentum density,
\begin{eqnarray}
     M^{\alpha\mu\nu} = && i \bar \psi \gamma^\alpha(x^\mu\partial^\nu
        -x^\nu\partial^\mu) \psi
      +  {i\over 4} \bar \psi\gamma^\alpha
      [\gamma^\mu,\gamma^\nu]\psi
    \nonumber \\  &&
      - F^{\alpha\sigma}(x^\mu\partial^\nu-x^\nu\partial^\mu)A_\sigma
     - F^{\alpha\mu}A^\nu + F^{\alpha\nu}A^\mu
      \ ,
\end{eqnarray}
where color indices are implicit. According to definition,
a nucleon moving in the $z$ direction
with momentum $P^\mu$ helicity 1/2 satisfies,
\begin{equation}
           J^{12}|P+\rangle = {1\over 2} |P+\rangle \ .
\end{equation}
Thus one can write down a spin sum rule,
\begin{eqnarray}
      {1\over 2} & = &\langle P+|J^{12}|P+ \rangle /\langle P+|P+ \rangle
         \nonumber \\
                 & = &{1\over 2}\Delta\Sigma + \Delta g  + L_q + L_g \ ,
\end{eqnarray}
where the matrix elements are defined as ($\gamma^5=i\gamma^0\gamma^1\gamma^2
\gamma^3$),
\begin{eqnarray}
         \Delta \Sigma & = & \langle P+|\hat S_{3q} |P+\rangle =
       \langle P+| \int d^3x \ \bar \psi \gamma^3\gamma_5\psi
     \ |P+\rangle \ , \nonumber \\
           \Delta g & = & \langle P+|\hat S_{3g} |P+\rangle
    =\langle P+| \int d^3x \ (E^1A^2-E^2A^1) \   |P+\rangle \ , \nonumber \\
           L_q & = & \langle P+|\hat L_{3q} |P+\rangle
   = \langle P+ | \int d^3x \ i\bar \psi
   \gamma^0(x^1\partial^2-x^2\partial^1) \psi \  |P+\rangle \ , \nonumber \\
           L_g & = &\langle P+|\hat L_{3g}|P+\rangle
   = \langle P+|\int d^3x \ E^i(x^2\partial^1-x^1\partial^2)A^i\ |P+\rangle \ ,
\end{eqnarray}
where for simplicity we have neglected the normalization of the
state.
It is clear from the above that $\Delta \Sigma$ and $\Delta g$
are the quark and gluon helicity contributions to the nucleon spin, and
$L_q$ and $L_g$ are the quark and gluon orbital
angular momentum contributions. Apart from
$\hat S_{3q}$, the other three operators $\hat S_{3g}$, $\hat L_{3q}$
and $\hat L_{3g}$ are not manifestly gauge invariant, and
thus a decomposition of the nucleon spin is in general
gauge-dependent. Furthermore, the matrix elements
depend on the choice of Lorentz frame. Only in light-front
coordinates and light-front gauge \cite{bb} $\Delta g$
is the gluon helicity measured
in high-energy scattering processes. We henceforth
work in this coordinates and gauge \cite{brodsky1} (the index
$0$ in Eq. (1) is now replaced by +).

The individual operators in $J^{12}$ are not
conserved charges, and hence
their matrix elements are generally divergent.
They can be renormalized in a scheme and the
renormalization introduces a scale dependence.
The scale dependence of $\Delta \Sigma$ and $\Delta g$
obeys the well-known Altarelli-Parisi (AP) equation, which
in the leading-log approximation is \cite{ap},
\begin{equation}
      {d\over dt} \left(\begin{array}{c}
                      \Delta \Sigma \\
                          \Delta g
                    \end{array} \right)
    = {\alpha_s(t)\over 2\pi} \left( \begin{array}{cc}
                       0 & 0 \\
                      {3\over 2}C_F & {\beta_0\over 2}
               \end{array} \right)
        \left(\begin{array}{c}
                      \Delta \Sigma \\
                          \Delta g
                    \end{array} \right) \ ,
\label{ape}
\end{equation}
where $t = \ln {Q^2/\Lambda_{\rm QCD}^2}$, $C_F=4/3$ and $\beta_0
=11-2n_f/3$ with $n_f$ the number of quark flavors. The
square matrix in the right-hand-side is called the splitting matrix.
In this paper, we derive an equation for
the leading-log evolution of the quark and gluon orbital
angular momenta. The result is
\begin{equation}
      {d\over dt} \left(\begin{array}{c}
                      L_q \\
                      L_g
                    \end{array} \right)
    = {\alpha_s(t)\over 2\pi} \left( \begin{array}{rr}
                       -{4\over 3}C_F & {n_f\over 3}\\
                      {4\over 3}C_F & -{n_f\over 3}
               \end{array} \right)
        \left(\begin{array}{c}
                      L_q \\
                      L_g
                    \end{array} \right) +
   {\alpha_s(t)\over 2\pi} \left( \begin{array}{rr}
                       -{2\over 3}C_F & {n_f\over 3}\\
                       -{5\over 6}C_F & -{11\over 2}
               \end{array} \right)
         \left(\begin{array}{c}
                      \Delta \Sigma \\
                          \Delta g
                    \end{array} \right) \ .
\label{hjt}
\end{equation}
We call the first term in the right hand side
the homogeneous term
and the second the inhomogeneous term (with corresponding splitting
matrices). The inhomogeneous term
was first studied by Ratcliffe \cite{ratcliffe}, from which
he concluded that orbital angular momentum plays an
essential role in the nucleon spin. At operator level, this
means that the orbital angular momentum
operators contain leading-twist contributions.
However, our result for the inhomogeneous part of
the $L_q-L_g$ evolution
disagrees with Ratcliffe's. Furthermore,
the homogenous term we have is new.

In the remainder of the paper, we sketch our derivation
of Eq. (\ref{hjt}) and find its general
solution. At the end of the paper, we go beyond
the leading-log approximation and discuss the
``anomaly cancellation" in the angular momentum operator $J^{12}$.
To keep the physics clear, we will follow
as closely as possible the original language of Altarelli
and Parisi in light-front coordinates although
the whole discussion can be made consistently in the language
of operator mixing. When necessary, we supplement
our discussion with the matrix elements of angular momentum
operators in composite parton states.

To begin, we review the standard derivation of the AP equation.
Consider a parent quark with momentum $p^\mu =(p^-=0, p^+,
 p_\perp=0)$ and
helicity $+1/2$, splitting into a daughter gluon of momentum
$k^\mu = (k^-, xp^+, k_\perp)$ and a daughter quark with momentum
$(p-k)^\mu$. Only 3-momentum is conserved during
the splitting. The total probability for the splitting is,
$\int^1_0 dx (1+(1-x)^2)/x$. [A multiplicative factor
$\alpha_s \ln Q^2/\mu^2$ is implied when we talk about
probability. $\mu^2$ here is an infrared cut-off which defines
the nature of the parent quark. It must be large enough so that
perturbative QCD is valid. $Q^2$ is an ultra-violet transverse
momentum cut-off which defines the scale of daughter partons.]
Since the quark helicity is conserved at the
leading-log, we therefore
have the item 0 in the upper-left corner of the AP
splitting matrix.
The helicity of the daughter gluon can either be $+1$ or $-1$.
The probabilities for both cases are $\int^1_0dx 1/x$ and
$\int^1_0dx  (1-x)^2/x$, respectively. The gluon
helicity produced in the splitting is just
$\int^1_0 dx (1-(1-x)^2)/x = (3/2)C_F$, which is the
element in the lower-left corner of the AP splitting matrix.

A further consideration of the above process leads
to an inhomogeneous source for the orbital
angular momenta.
Since the total angular momentum in the $z$ direction
is conserved in the splitting, orbital angular momentum
has to be produced to cancel the helicity of the
daughter gluon. This means that total orbital angular momentum
carried by the daughter quark
and gluon is $-(3/2)C_F$. QCD determines partition of this
between the quark and gluon according to the
matrix elements,
\begin{equation}
         \langle p+|\hat L_{3q}|p+ \rangle\ ,~~~~
          \langle p+|\hat L_{3g}| p+ \rangle ,
\end{equation}
where $|p+\rangle$ is the quark-gluon state
produced in the splitting. To calculate them,
we start with less singular off-forward matrix elements
and consider their forward limit\cite{jaffemanohar}.
We find,
\begin{eqnarray}
     \langle p+|\hat L_{3q}|p+\rangle
     & = &-  C_F {\cal N}\int^1_0 dx x \left({1\over x}
   - {(1-x)^2\over x}\right)  \nonumber \\
    & = & - {2\over 3}C_F {\cal N}\ , \nonumber \\
     \langle p+|\hat L_{3g}|p+\rangle
     & = & - C_F{\cal N}\int^1_0 dx (1-x) \left({1\over x}
   - {(1-x)^2\over x}\right)  \nonumber \\
        & = & - {5\over 6}C_F{\cal N}  \ ,
\end{eqnarray}
where ${\cal N} = 2p^+(2\pi)^3\delta^3(0)(\alpha_s/2\pi)\ln Q^2/\mu^2$.
The coefficients in front of ${\cal N}$
give the first column of the inhomogeneous splitting
matrix in Eq. (\ref{hjt}). Interestingly, the sharing of
the orbital angular momentum is
done according to the $x$ and $1-x$ moments of the polarized
gluon density in the parent quark.

A similar discussion leads to the second columns of the
AP splitting matrix and the
inhomogeneous splitting matrix in Eq. (\ref{hjt}).
Let us remark here the physical origin
for the well-known result that the gluon helicity
increases logarithmically in the asymptotic limit
\cite{ar}. When a gluon
splits, there are four possible final states: 1) a quark
with helicity 1/2 and an antiquark with helicity $-1/2$;
2) a quark with helicity $-1/2$ and an antiquark with
helicity $1/2$; 3) a gluon with helicity $+1$ and another
with helicity $-1$; 4) two gluons with helicity $+1$.
In the first two processes, there is a loss of the
gluon helicity with probability $(n_f/2)\int^1_0dx (x^2+(1-x)^2)$.
In the third process, there is also a loss of the gluon
helicity, with probability $\int^1_0dx(x^3/(1-x) + (1-x)^3/x)$.
In the last process, there is a gain of the gluon  helicity with
probability $\int^1_0 dx 1/(x(1-x))$. When summed,
the gluon helicity has a net gain with probability $11/2-n_f/3=\beta_0/2$
in the splitting. Thus at increasingly smaller
distance scales, gluons split consecutively and the
helicity builds up logarithmically.

Now we turn to the homogenous term in the orbital
angular momentum evolution equation. To calculate the splitting
matrix, we consider the matrix elements of
$\hat L_{3q}$ and $\hat L_{3g}$ in a
parton state with non-vanishing transverse momentum.
In a bare parton state, we have,
\begin{equation}
   \langle p'+|\hat L_{3q}|p+\rangle = 2p^+ (2\pi)^3\left(-ip_1' {\partial\over
        \partial p_2'} + ip_2' {\partial\over
        \partial p_1'}\right)\delta^3(p'-p) \ .
\label{bm}
\end{equation}
The derivative on the $\delta$-function
means that when the distribution is
convoluted with a test function, the derivative
will be taken of the test function. When
calculating the matrix element in the
composite parton states from the parton splitting, we have
\begin{equation}
     \langle p'+|\hat L_{3q}|p +  \rangle
       = 2p^+(2\pi)^3  \phi(p',p) \left( -ip_1' {\partial\over
        \partial p_2'} + ip_2' {\partial\over
        \partial p_1'}\right)\delta^3(p'-p) \ .
\end{equation}
When the distribution is convoluted with a test function, the derivative
will be taken of $\phi(p',p)$ and then of
the test function. The first term represents
generation of the orbital angular momentum
from the parton helicity discussed previously.
The second term has the same
structure as the basic matrix element in Eq. (\ref{bm})
and represents the self-generation of
orbital angular momentum in the splitting.

According to the above recipe, we
calculate the gluon orbital angular momentum
generated from the quark orbital angular momentum,
\begin{eqnarray}
      \phi(p,p)
    & = & {\alpha_s \over 2\pi} \ln{Q^2\over \mu^2}
       C_F \int^1_0 x{(1+(1-x)^2))\over x} dx  \nonumber \\
    & = & {\alpha_s \over 2\pi} \ln{Q^2\over \mu^2} C_F {4\over 3} \ .
\end{eqnarray}
Interestingly, this is the same as the fraction of the
quark momentum carried by the daughter gluon in the infinite
momentum frame. The net loss of the quark
orbital angular momentum in the
splitting must be the negative
of the above. These numbers form
the first column of the homogeneous splitting matrix.
A similar calculation yields the second column.

The solution of the evolution equation can be obtained
straightforwardly. First, let us write down the well-known
result for the quark and gluon helicities\cite{ratcliffe,ar},
\begin{eqnarray}
       \Delta \Sigma(t) & = & {\rm const} \ , \nonumber  \\
       \Delta g(t) & = & -{4\Delta \Sigma \over \beta_0} + {t\over t_0}
            \left(\Delta g_0 + {4\Delta \Sigma\over \beta_0}\right)\ .
\label{sol1}
\end{eqnarray}
The second equation exhibits the famous behavior of the
gluon helicity: increasing like $\ln Q^2$ as $Q^2\rightarrow \infty$.
The coefficient of the term depends on the special combination
of initial quark and gluon helicities, which is likely positive
at low-momentum scales
according to the recent experimental data on $\Delta \Sigma $\cite{e143}
and theoretical estimates for $\Delta g$ \cite{brodsky,jaffe}.
The solution for the orbital momenta is,
\begin{eqnarray}
      L_q(t) & = & -{1\over 2}\Delta \Sigma + {1\over 2}{3n_f\over
          16+3n_f} +
           {(t/ t_0)}^{-{2(16+3n_f)/ (9\beta_0)}}
            \left(L_q(0) + {1\over 2}\Delta \Sigma -
            { 1\over 2}{3n_f\over 16+3n_f}\right)\ , \nonumber \\
      L_g(t) & = & -\Delta g(t) + {1\over 2}{16\over
          16+3n_f} +
           {(t/t_0)}^{-{2(16+3n_f)/(9\beta_0)}}
            \left(L_g(0) + \Delta g(0) -
           { 1\over 2}{16\over 16+3n_f}\right)\ .
\end{eqnarray}
Given a composition of the nucleon spin at some initial scale
$Q_0^2$, the above equations yield the spin composition at
any other perturbative scales in the leading-log approximation.
 From the expression for $L_g(t)$, it is clear that that the large
gluon helicity at large $Q^2$ is cancelled by an
equally large, but negative, gluon orbital angular momentum.

Neglecting the sub-leading terms at large $Q^2$,
we get,
\begin{eqnarray}
      J_q &=& L_q + {1\over 2}\Delta \Sigma = {1\over 2}{
      3n_f\over 16 + 3n_f}\ ,  \nonumber \\
      J_g &=& L_g + \Delta g = {1\over 2}{
      16\over 16 + 3n_f}\ .
\end{eqnarray}
Thus partition of the nucleon spin between quarks
and gluons follows the well-known partition of the nucleon momentum
\cite{gw} !
Mathematically, one can understand
this from the expression for the QCD angular momentum
density $M^{\mu\alpha\beta} = T^{\mu\alpha}x^\beta
     - T^{\mu\beta} x^\alpha$. When $M^{\mu\alpha\beta}$
and $T^{\alpha\beta}$ are each separated
into gluon and quark contributions, the anomalous
dimensions of the corresponding terms are the same
because they have the
same short distance behavior.

It is interesting to speculate phenomenological
consequences of this asymptotic partition of the nucleon spin.
Assuming, as found in the case of the
momentum sum rule\cite{emc},
that the evolution in $Q^2$ is very slow, then
the above partition may still be roughly correct
at low momentum scales, say, $Q^2\sim 3$ GeV$^2$. If
this is the case, from the experimentally measured
$\Delta \Sigma$ we get an estimate of the
quark orbital contribution at these scales,
\begin{equation}
       L_q = 0.05 \sim 0.15 \ .
\end{equation}
To find a separation of the gluon contribution into
spin and orbit parts, we need to know $\Delta g$,
which shall be measurable in the future\cite{rsc}.
However, if $Q^2$ variation is rapid, the asymptotic
result implies nothing about the low $Q^2$ spin
structure of the nucleon. Unfortunately, no one knows yet
how to measure $L_q$ to determine the role of
$Q^2$ variation.

So far we have considered the leading-log results. It
is well-known that at the next-to-leading-log level,
the matrix element of the axial current in a gluon state
contains an anomaly \cite{ar,ccm}. Because of this anomaly, the axial
current acquires an anomalous
dimension starting at the two-loop level \cite{ar,kodaira,kaplan}.
Also because of the anomaly, it was suggested that
the quark helicity for light flavors be modified to
subtract the anomaly contribution\cite{et,ar,ccm}.
Since the axial current appears in the
conserved charge $J^{12}$, the anomaly
contribution to $\Delta \Sigma$ must be cancelled by
matrix elements of some other operators in $J^{12}$.
Below, we show such cancellation comes from the
quark orbital angular momentum $\hat L_{3q}$.

To investigate the anomaly cancellation, we consider
a gluon moving in the $z$ direction
with momentum $p^\mu$ and helicity $+1$.
The gluon can virtually
split into a quark-antiquark pair.
For zero quark masses, the quark and antiquark
have opposite helicities due to chirality conservation.
Thus the helicity of the gluon is entirely
transferred to the orbital angular momentum of
the pair. However, in the limit of the quark transverse momentum
going to infinity, the chiral symmetry
is broken by the anomaly. As a consequence,
the quark and antiquark can have same helicity.
Indeed, the calculation of Carlitz, Collins and Mueller
shows \cite{ccm},
\begin{equation}
    \langle p+|{1\over 2}\bar \psi \gamma^+\gamma_5
 \psi| p+\rangle
     = - {\alpha_s\over 4\pi} \times 2p^+ \  .
\end{equation}
The sign indicates that the quark and antiquark
prefer to have helicity $-1/2$. To preserve the total angular momentum,
the orbital motion of the pair has the same probability
to carry two units of angular momentum. Indeed, an explicit
calculation yields,
\begin{equation}
   \langle p+|\hat L_{3q}|p+\rangle  - \langle p+|p+\rangle
     =  {\alpha_s\over 4\pi} \times 2p^+ (2\pi)^3\delta^3(0) \ .
\end{equation}
According to the above, if one redefines the quark helicity
by subtracting the anomalous gluon contribution\cite{ar,ccm},
one must also redefine the quark orbital angular momentum
by adding to it the same amount. Thus if a large $\Delta g$ helps
to restore the measured $\Delta \Sigma$ to the quark model result, the
same effect will make $L_q$ in Eq. (16) negative.
However, the fraction of the nucleon
spin carried by quarks is invariant under such redefinition.

To summarize, we have derived an evolution equation for the
quark and gluon orbital angular momenta in QCD. In the asymptotic
limit, the solution of the equation indicates that the
quark and gluon contributions to the nucleon spin
are the same as their contributions to
the nucleon momentum. Furthermore, both the
gluon orbital and helicity contributions grow
logarithmically at large $ Q^2$ and with opposite sign,
so their sum stays finite. We show that the anomaly in
the quark helicity is cancelled by a contribution
in the quark orbital angular momentum, so the total
quark contribution to the nucleon spin is anomaly-free.
We also discussed phenomenological implications
of our results, in particular, about the size
of the quark orbital angular momentum.

\acknowledgements
We wish to thank I. Balitsky for
useful conversations, and P. Ratcliffe and A. Sch\"afer
for helpful comments. PH would like to acknowledge
the hospitality accorded to him by the Center for Theoretical
Physics.


\begin{references}
\frenchspacing

\bibitem{ji}
X. Ji, Phys. Rev. Lett. 74 (1995) 1071; Phys. Rev. D52 (1995) 271.

\bibitem{jaffemanohar}
R. L. Jaffe and A. Manohar, Nucl. Phys. B337 (1990) 509.

\bibitem{bb}
I. Balitsky and V. Braun, Phys. Lett. B267 (1991) 405.

\bibitem{brodsky1}
For a nice introduction to QCD in light-front coordinates, see
the appendix of the paper by S. Brodsky and P. Lepage,
in {\it Perturbative Quantum Chromodynamics},
ed. by A. Mueller, World Scientific, Singapore, 1989.

\bibitem{ap}
G. Altarelli and G. Parisi, Nucl. Phys. B126 (298).

\bibitem{ratcliffe}
P. G. Ratcliffe, Phys. Lett. B192 (1987) 180.

\bibitem{ar}
G. Altarelli and G. G. Ross, Phys. Lett. B212 (1988) 391.

\bibitem{e143}
J. Ashman et al., Nucl. Phys. B 328 (1989) 1;
B. Adeva et al., Phys. Lett. B302 (1993) 533;
P. L. Anthony et al., Phys. Rev. Lett. 71 (1993) 959;
K. Abe et al., Phys. Rev. Lett. 75 (1995) 25.

\bibitem{brodsky}
S. J. Brodsky, M. Burkardt, I. Schmidt, Nucl. Phys. B441 (1995) 197.

\bibitem{jaffe}
R. L. Jaffe, MIT-CTP-2466, HUTP-95/A034, hep-ph/9509279.

\bibitem{gw}
D. Gross and F. Wilczek, Phys. Rev. D9 (1974) 980.

\bibitem{emc}
See, for instance, T. Sloan, G. Smadja, and R. Voss, Phys. Rep. 162 (1988) 45.

\bibitem{rsc}
G. Bunce et al., Particle World 3 (1992) 1; The STAR and
PHENIX spin proposals.

\bibitem{ccm}
R. D. Carlitz, J. C. Collins, and A. H. Mueller,
Phys. Lett. B214 (1988) 229.

\bibitem{kaplan}
D. B. Kaplan and A. Manohar, Nucl. Phys. B310 (1988) 527.

\bibitem{kodaira}
J. Kodaira, Nucl. Phys. B 165 (1979) 129.

\bibitem{et}
A. V. Efremov and O. V. Teryaev, JINR Report E2-88-287 (1988).

\nonfrenchspacing
\end{references}
\end{document}